\documentclass[seceq]{ptptex}

\usepackage{graphicx,color}

\newcommand\CL{{\mathcal L}}

\newcommand\CH{{\mathcal H}}

\renewcommand\NG{\mathrm{NG}}
\newcommand\WZ{\mathrm{WZ}}
\newcommand\e{\mathrm{e}}

\title{
A Covariant Approach to Noncommutative M5-branes
\footnote{A contribution given by K.~Y.\ to Nishinomiya-Yukawa Memorial
Symposium on Theoretical Physics ``Noncommutative Geometry and Quantum
Spacetime in Physics'' (Nov.\ 11--15, 2006, Japan).
}  }

\author{
Makoto \textsc{SAKAGUCHI}$^{1,}$\footnote{
makoto$\_$sakaguchi@pref.okayama.jp}    
and Kentaroh \textsc{YOSHIDA}$^{2,}$\footnote{kyoshida@post.kek.jp}
}

\inst{
$^{1}$Okayama Institute for Quantum Physics (OIQP), \\ 
1-9-1 Kyoyama, Okayama 700-0015, Japan. \\
 $^{2}$Theory Division, Institute of Particle and Nuclear Studies, \\
High Energy Accelerator Research
Organization (KEK), \\ 1-1 Oho, Tsukuba, Ibaraki 305-0801, Japan. 
}

\abst{
We briefly review how to discuss noncommutative (NC) M5-branes and
intersecting NC M5-branes from $\kappa$-invariance of an open
supermembrane action with constant three-form fluxes. The
$\kappa$-invariance gives rise to possible 
Dirichlet brane configurations. 
We shortly summarize a construction of projection operators for NC
M5-branes and some intersecting configurations of NC M5-branes. A strong
flux limit of them is also discussed. 
\vspace*{0.2cm} \\
\rightline{KEK-TH-1139, OIQP-07-03, {\tt hep-th/0702132}} 
}

\begin{document}

\maketitle

\section{Introduction}

Supermembrane theory in eleven dimensions \cite{BST,dWHN} is closely
related to the M-theory formulation \cite{BFSS}, where open membranes
\cite{Strominger,Townsend} can be considered as well as closed ones.
Open membranes can end on a $p$-dimensional Dirichlet $p$-brane for
$p=1,5$ and 9 \cite{EMM,dWPP} just like an open string can attach to
D-branes. The $p=5$ case corresponds to M5-brane and the $p=9$ is the
end-of-world 9-brane in the Horava-Witten theory \cite{HW}. 
 
The Dirichlet branes can be investigated from the $\kappa$-symmetry
argument \cite{EMM,LW}. It is a covariant way and a specific
gauge-fixing such as light-cone gauge is not necessary. Then it is
sufficient to consider a single action of open string or open membrane,
rather than each of D-brane actions. It is moreover easy to find what
configurations are allowed to exist for rather complicated D-brane
setups such as intersecting D-branes or less supersymmetric D-branes,
which are difficult to discuss within a brane probe analysis. The method
is not restricted to a flat spacetime and can be generalized to curved
backgrounds \cite{AdS-memb}. 

\section{The $\kappa$-symmetry Argument}
  
The Green-Schwarz action of a supermembrane in flat spacetime is
composed of the Nambu-Goto (NG) part and the Wess-Zumino (WZ) part
\cite{BST}
\begin{eqnarray}
S&=&\int_\Sigma d^3\sigma\left[
\CL_{\NG}
+\CL_{\WZ}
\right]\,.  \nonumber 
\end{eqnarray}
Since the bulk action is $\kappa$-invariant, the $\kappa$-variation of
the action $\delta_\kappa S$ leaves only surface terms.  The NG part
does not give rise to any surface terms. Thus it is sufficient to
examine the $\kappa$-variation of the WZ part,
\begin{eqnarray}
\delta_\kappa S_{\WZ}|&=&\int_{\partial\Sigma} d ^2\xi
\left[
\CL^{(2)}+\CL^{(4)}+\CL^{(6)}
\right]\,, \nonumber \\
\CL^{(2)}&=&-i
\left[
\bar\theta\Gamma_{\bar A\bar B}\delta_\kappa\theta
+\CH_{\bar A\bar B\bar C}\bar\theta\Gamma^{\bar C}\delta_\kappa\theta
\right]\dot X^{\bar{A}}{X'}^{\bar B}~,
\label{second order in theta}
\\
\CL^{(4)}&=&
\Big[
-\frac{3}{2}\bar\theta\Gamma^{ A}\delta_\kappa\theta~
\bar\theta\Gamma_{ A\bar B}
+\frac{1}{2}\bar\theta\Gamma_{ A\bar B}\delta_\kappa\theta~
\bar\theta\Gamma^{ A}
\Big](\theta'\dot X^{\bar B}-\dot \theta {X'}^{\bar B})\,, 
\label{surface term theta^4}
\end{eqnarray}
where the sixth order part $\CL^{(6)}$ disappears due to the Fierz
identity. Here we have already utilized bosonic boundary conditions
\cite{Bergshoeff}. In order to ensure the $\kappa$-invariance these
surface terms should vanish. Thus the problem of finding possible
Dirichlet branes is boiled down to constructing the projection operators
to make (\ref{second order in theta}) and (\ref{surface term theta^4})
vanish. It can be performed by constructing a gluing matrix $M$
satisfying $\theta = M\theta$ on the boundary.

\section{Noncommutative M5-brane} 

A single NC M5-brane (012345) with $\CH_{012}$ and $\CH^{345}$ is
characterized, for example, by the gluing matrix\cite{SY:NCM},
\begin{align}&
M=h_0\Gamma^{012345}
+h_1\Gamma^{012}\,.
\label{M-A}
\end{align}
For $M$ to define a projection, $M^2=1$ should be satisfied. Then we
obtain the following condition,
\begin{eqnarray}
h_0^2 + h_1^2=1\,. 
\label{M5+M2: M^2=1}
\end{eqnarray}
We can see that (\ref{second order in theta}) may vanish by imposing the
conditions
\begin{eqnarray}
h_1-\CH_{012}=0~,~~~
h_1-h_0\CH^{345}=0~.
\label{solution M5}
\end{eqnarray}
It is easy to see that (\ref{surface term theta^4}) also becomes zero,
and the gluing matrix (\ref{M-A}) with the two conditions (\ref{M5+M2:
M^2=1}) and (\ref{solution M5}) gives a possible M5-brane configuration.

Then let us consider the interpretation of the solution constructed
above. By substituting (\ref{solution M5}) for (\ref{M5+M2: M^2=1}), we
obtain the following condition,
\begin{eqnarray}
\frac{1}{(\CH^{345})^2}
-\frac{1}{(\CH_{012})^2}
=-1\,. \nonumber 
\end{eqnarray}
This is nothing but the self-dual condition \cite{Sezgin} of the gauge
field on the M5-brane \cite{SW}. That is, we have reproduced the
information on the NC M5-brane from the $\kappa$-symmetry argument for
the open supermembrane action. Thus we recognize that the projection
operator should describe the NC M5-brane.

Let us consider a commutative limit and a strong flux limit. The
conditions (\ref{M5+M2: M^2=1}) and (\ref{solution M5}) are solved by
using an angle variable $\varphi$\,,
\begin{eqnarray}
&&h_0=\cos\varphi\,, \quad h_1=\sin\varphi\,, \quad 
\CH_{012}=\sin\varphi\,, \quad 
\CH^{345}=\tan\varphi \qquad (0\le \varphi \le\pi/2)\,. 
\nonumber 
\end{eqnarray}
Then the gluing matrix $M$ is written as
\begin{eqnarray}
M=\e^{\varphi\Gamma^{345}}\Gamma^{012345}\,. \nonumber 
\end{eqnarray}
For a commutative limit $\varphi\to 0$, the NC M5 reduces to commutative
M5 (012345), since $\CH\to0$ and $M\to \Gamma^{012345}$.

On the other hand, for $\varphi\to\pi/2$\,, we see that
$\CH^{345}\to\infty$ and so the gluing condition reduces to
$M\to\Gamma^{012}$ with a critical flux $\CH_{012}=1$\,. It seems that
the resulting projection operator should describe a critical M2-brane
(012). Eventually this limit is nothing but the OM limit \cite{OM} and
it should correspond to one of infinitely many M2-branes dissolved on
the M5-brane.  This is analogous to the D2-D0 system where a D2-brane
with a flux reduces to a D2-brane with infinitely many D0-brane in a
strong magnetic flux limit.

As is well known, the $p=2$ case is not allowed as a projection operator
in the case without fluxes. Hence it is a non-trivial problem whether
the resulting projection operator for a critical M2 is consistent to the
$\kappa$-symmetry. The $p=2$ case is actually special among other $p$\,,
and due to some identities intrinsic to $p=2$\,, (\ref{second order in
theta}) vanishes when 
\begin{eqnarray}
\CH_{012}=1~.  \label{fixed H for NC M2 }
\end{eqnarray}
It is easy to show that (\ref{surface term theta^4}) disappears. Thus we
have checked that the $\kappa$-variation surface terms should vanish for
an M2-brane with the critical $\CH$ (\ref{fixed H for NC M2 }). Although
the $\kappa$-symmetry is maintained for the M2-brane, the charge
conservation\cite{Strominger} requires the existence of M5-brane behind
M2-branes. That is, a NC M5-brane should be regarded as a bound state of
M5 and M2.

\section{Intersecting Noncommutative M5-branes}

In comparison to the case of a single NC M5-brane, a configuration of
intersecting NC M5-branes is characterized by two gluing
matrices\cite{SY:int},
\begin{eqnarray}
M_1=\e^{\varphi_1\Gamma^{A_0A_1A_2}}\Gamma^{A_0\cdots A_5}\,, \quad
M_2=\e^{\varphi_2\Gamma^{B_0B_1B_2}}\Gamma^{B_0\cdots B_5}\,,\quad 
[M_1,M_2]=0\,. \nonumber 
\end{eqnarray}
The requirement $[M_1,M_2]=0$ leads to the four possibilities for the
projection. As an example, let us focus upon one of these cases, NC
M5$\bot$NC M5(3) described by 
\begin{eqnarray}
&& M_1=\e^{\varphi_1\Gamma^{235}}\Gamma^{012345}~,~~~
\CH_{014}=\sin\varphi_1~,~~~
\CH^{235}=\tan\varphi_1 \qquad (0\le \varphi_{1}\le\pi/2)\,, 
\nonumber \\ 
&& M_2=\e^{\varphi_2\Gamma^{137}}\Gamma^{012367}~,~~~
\CH_{026}=-\sin\varphi_2~,~~~
\CH^{137}=\tan\varphi_2 \qquad (0\le \varphi_{2}\le\pi/2)\,. \nonumber 
\end{eqnarray}
It reduces to a commutative M5 (012345)$\bot$M5
(012367)\cite{PT,Tseytlin,GKT1} in the limit $\varphi_{1,2} \to 0$\,.

\begin{figure}[htb]
  \begin{center}
    \includegraphics[keepaspectratio=true,height=59mm]{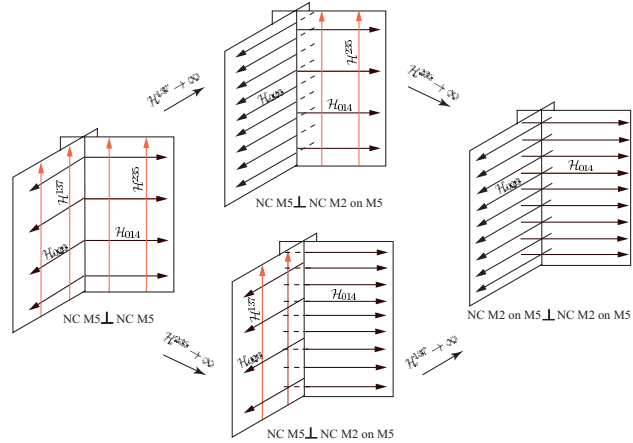}
  \end{center}
  \caption{\footnotesize 
Two sequences of the strong flux limits of NC M5$\bot$NC M5 (3)}
  \label{fig:NCM5M5c.eps}
\end{figure}

M2$\bot$M5 (1)\cite{PT,GKT1} can be realized from the NC M5$\bot$NC M5
(3) by taking a strong flux limit. In the limit $\varphi_2\to\pi/2$\,,
we obtain a NC version of M2$\bot$M5 (1),
\begin{eqnarray}
&& M_1=\e^{\varphi_1\Gamma^{235}}\Gamma^{012345}\,, \qquad 
M_2=-\Gamma^{026}\,. \nonumber 
\end{eqnarray}
Further letting $\varphi_1\to\pi/2$, we obtain M2 (014)$\bot$M2
(026)\cite{PT,GKT1}. The other way is possible and the two sequences of
the strong flux limits are depicted in Fig.\ \ref{fig:NCM5M5c.eps}. It
is also possible to discuss NC M5$\bot$C M5 (1) \cite{SY:NCM}.

\subsubsection*{Acknowledgements} 
This work is supported in part by the Grant-in-Aid for Scientific
Research (No.~17540262 and No.~17540091) from the Ministry of Education,
Science and Culture, Japan.  The work of K.~Y.\ is supported in part by
JSPS Research Fellowships for Young Scientists.

\end{document}